\documentclass[]{spie}  

\usepackage{amsmath,amsfonts,amssymb}
\usepackage{graphicx}
\usepackage[colorlinks=true, allcolors=blue]{hyperref}
\usepackage{subcaption}
\usepackage{float}

\title{Lessons learned from the NEAR experiment and prospects for the upcoming mid-IR HCI instruments}

\author[a]{Prashant Pathak}
\author[b]{Markus Kasper}
\author[a]{Olivier Absil} 
\author[a]{Gilles Orban de Xivry} 
\author[b]{Ulli Käufl}
\author[b]{Gerd Jakob}
\author[b]{Ralf Siebenmorgen}
\author[b]{Serban Leveratto}
\author[c]{Eric Pantin}

\affil[a]{Space sciences, Technologies and Astrophysics Research (STAR) Institute, Universit\'e de Li\`ege, 19c All\'ee du Six Ao\^ut, 4000 Li\`ege, Belgium}
\affil[b]{European Southern Observatory, Karl-Schwarzschild-Str. 2, 85748 Garching bei München, Germany}
\affil[c]{AIM, CEA, CNRS, Université Paris-Saclay, Université Paris Diderot, Sorbonne Paris Cité, F-91191 Gif-sur-Yvette, France}

\authorinfo{Further author information: (Send correspondence to.)\\E-mail: ppathak8@gmail.com, Telephone: +32 497039597}

\begin{document} 
\maketitle

\begin{abstract}
The mid-infrared (IR) regime is well suited to directly detect the thermal signatures of exoplanets in our solar neighborhood. The NEAR experiment: demonstration of high-contrast imaging (HCI) capability at ten microns, can reach sub-mJy detection sensitivity in a few hours of observation time, which is sufficient to detect a few Jupiter mass planets in nearby systems. One of the big limitations for HCI in the mid-IR is thermal sky-background. In this work, we show that precipitate water vapor (PWV) is the principal contributor to thermal sky background and science PSF quality. In the presence of high PWV, the HCI performance is significantly degraded in the background limited regime. 
\end{abstract}


\section{INTRODUCTION} \label{sec:intro}  
The direct imaging of habitable exoplanets is challenging due to angular resolution and high-contrast requirements. The field of high-contrast imaging (HCI) is rapidly advancing to achieve such a goal.
Most of the current HCI instruments operate in the near-infrared (IR) regime. 

The near-IR regime is more sensitive to self-luminous planets and the mid-IR (8-13$\mu$m) is more sensitive to colder planets and less massive planets~\cite{quanz2015}. The big limitations for HCI in the mid-IR are reduced angular resolution and large thermal sky-background for ground-based observations. Therefore, the mid-IR is best suited to look for exoplanets around nearby stars. 

The NEAR (New Earths in the Alpha Cen Region) experiment was an outcome of a collaboration between Breakthrough and ESO (European Southern Observatory). The goal of the NEAR  experiment is to enable HCI capability at $10~\mu m$ and to look for low mass exoplanets in the $\alpha$~Cen A/B  binary system~\cite{Kasper2017, Kasper2019}. The project involved upgrading the existing VISIR (Very Large Telescope Imager and Spectrometer for the mid-InfraRed) instrument~\cite{lagage2004} with Shack–Hartmann based AO system ($>95\%$~Strehl ratio in the science band) and high-performance coronagraph such as annular groove phase mask (AGPM) at $10~\mu m$.

The NEAR was able to reach a final contrast of $\approx 3 \times 10^{-6}$ at $1''$ ($3.5~\lambda$/D) sufficient for the detection of Neptune mass planets in the habitable zone of $\alpha$ Cen A. A candidate with an SNR of 3 was found whose nature (e.g., planet, part of a zodiacal disk, image artifact) remains to be confirmed by follow-up observations~\cite{Wagner2021}. Recently, using the Keplerian-stacker algorithm same signal was confirmed with a higher SNR of 5~\cite{kstacker2022}. The NEAR was also offered for observations under ESO's science demonstration program. One such program involved searching for exoplanets around $\epsilon$ Indi A, $\epsilon$ Eri, $\tau$ Ceti, Sirius A, and Sirius B. No new planets were found, but a new upper limit was found for all the targets in direct imaging~\cite{Pathak2021}.

In this work, we use $96.2$~hrs. of data collected under $\alpha$~Cen campaign to understand HCI performance affected in the presence of atmospheric parameters and instrumental limitations. 

In section~\ref{sec:observation} we describe the observation and data reduction techniques. In section~\ref{sec:results} we describe the results for various analyses. 

\begin{table}[ht]
\caption{Observing parameters for all the nights under various atmospheric conditions. All the observations were carried out using the NEAR N-band filter with a bandpass of $10-12.5~\mu m$.} 
\label{tab:obs}
\begin{center}       
\begin{tabular}{|c|c|c|c|c|c|} 
\hline
\rule[-1ex]{0pt}{3.5ex}  Observation & Time & Median   & Median & Median & Median \\
\rule[-1ex]{0pt}{3.5ex}   night      & (hr) & PWV (mm) & seeing ('') & Temp ($^{\circ}$C) & RH ($\%$)\\
\hline
\rule[-1ex]{0pt}{3.5ex}  23/05/2019 & 7.64 & 0.32  & 0.9     & 	11.9 & 	4 \\
\rule[-1ex]{0pt}{3.5ex}  25/05/2019 & 7.19 & 3.83  & 1.1     & 	8.1  & 	9\\
\rule[-1ex]{0pt}{3.5ex}  27/05/2019 & 3.00 & 5.45  & 1.3	 &  6.8  & 9\\
\rule[-1ex]{0pt}{3.5ex}  29/05/2019 & 7.26 & 9.10  & 0.7	 &  8.9	 & 24\\
\rule[-1ex]{0pt}{3.5ex}  30/05/2019 & 6.78 & 3.83  & 0.9	 &  7.3  & 	36\\
\rule[-1ex]{0pt}{3.5ex}  31/05/2019 & 6.50 & 3.51  & 0.8	 &  7.2	 & 16\\
\rule[-1ex]{0pt}{3.5ex}  01/06/2019 & 5.58 & 0.51  & 0.8	 &  9.1	 & 14.5\\
\rule[-1ex]{0pt}{3.5ex}  02/06/2019 & 7.98 & 2.57  & 0.7	 &  14.1 & 	4.5\\
\rule[-1ex]{0pt}{3.5ex}  03/06/2019 & 7.81 & 3.83  & 0.7	 &  15.8 & 	5\\
\rule[-1ex]{0pt}{3.5ex}  04/06/2019 & 6.67 & 2.23  & 1.0	 &  16.4 & 8 \\
\rule[-1ex]{0pt}{3.5ex}  05/06/2019 & 4.00 & 1.71  & 1.1	 &  15.7 & 	3.5\\
\rule[-1ex]{0pt}{3.5ex}  07/06/2019 & 3.20 & 1.33  & 0.5	 &  14.0 & 	3\\
\rule[-1ex]{0pt}{3.5ex}  08/06/2019 & 7.21 & 1.81  & 0.7	 &  14.0 & 3.5\\
\rule[-1ex]{0pt}{3.5ex}  09/06/2019 & 3.84 & 2.24  & 0.5     & 	14.9 & 	4\\
\rule[-1ex]{0pt}{3.5ex}  10/06/2019 & 2.69 & 2.91  & 1.4	 &  12.8 & 8.5\\
\rule[-1ex]{0pt}{3.5ex}  26/06/2019 & 6.10 & 5.45  & 0.9	 &  10.2 & 	17\\

\hline 
\end{tabular}
\end{center}
\end{table}
\begin{figure}[ht]
	\begin{center}
		\begin{tabular}{c} 
			\includegraphics[width=\textwidth]{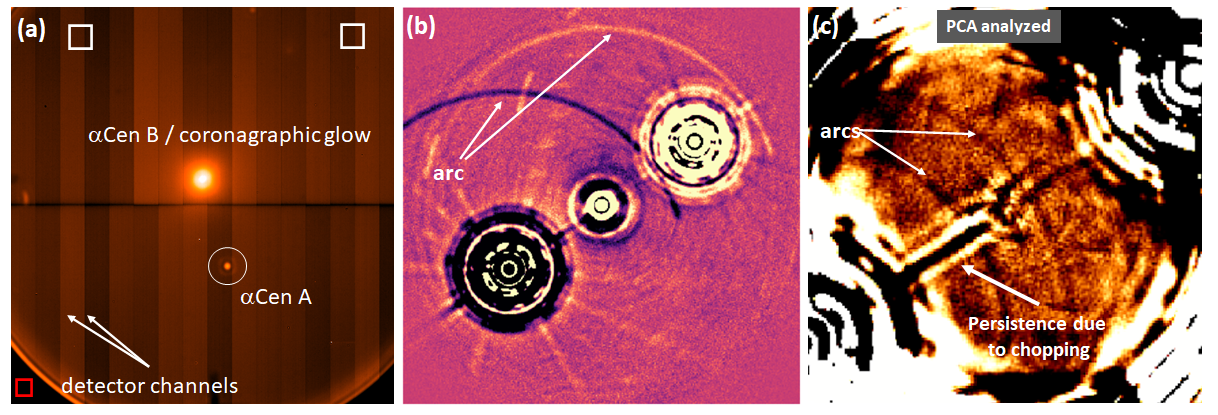}
		\end{tabular}
	\end{center}
	\caption[example] 
	{ \label{fig:det} 
		(a) A single exposure raw science frame. (b) A derotated median filtered averaged science frame for one night of observation. (c) Same as (b) with PCA analysis.}
\end{figure} 

\section{OBSERVATIONS AND DATA REDUCTION}\label{sec:observation}
The planned NEAR campaign ran from $23^{rd}$ May until $11^{th}$ June of 2019, with a total of 20 nights accounting for 90.2 hours of observations. Several nights were lost due to weather, to compensate for the loss, an additional night of observation was carried out on $26^{th}$ of June 2019. The final observation time was 96.2 hrs. For the analysis presented in this work, we removed 5 nights with less than 1.7 hrs of observation time, this resulted in 16 nights with 93.35 hrs to work with. The summary of the final selected nights with various atmospheric conditions is summarized in Table~\ref*{tab:obs}.

\subsection{Observation}
The $\alpha$ Cen campaign observations incorporate a common strategy for all the observing nights, these include the use of AO, a high-performance annular groove phase mask (AGPM) coronagraph, and chopping. The chopping was performed by the deformable secondary mirror (DSM) of the VLT and centering of the on-axis targets on the coronagraphs with the help of the QACITS algorithm~\cite{qacits2020}. The chop throw of the DSM was $~4.9''$, which sufficient to chop between $\alpha$~Cen A and B. The chopping was performed (typical speed of $10~Hz$) to reduce the sky background and the excess low-frequency noise common to the mid-IR arrays (Si:As array)~\cite{Arrington1998}. 

A uniform exposure of $5.992$~msec was used for all the nights except a part of 1st night $23{rd}$~May. 
A two-hour of observation was performed with an exposure of $5.493$~msec. To keep the data uniform for all the analyses, 2 hrs of observation with smaller exposure time is excluded. 

\subsection{Thermal sky background calculation}
For calculating thermal sky background, individual exposure raw frames were used. An example of the raw frame is shown in Figure~\ref{fig:det}~(a). The background flux was measured by calculating the total flux per pixel on the region of frames free from any stellar residuals (at the top two corners) using small square boxes as shown in Figure~\ref{fig:det}~(a). Next, an average was calculated by dividing by the area to estimate background values per pixel. The detector bias was estimated by calculating the flux inside the red box and dividing by area as shown in Figure~\ref{fig:det}~(a). The sky background was calculated by subtracting the detector bias from the total flux. 

The values of airmass, seeing, temperature, PWV, and relative humidity were extracted from the header of each fits files. Apart from the airmass, values of atmospheric parameters are measured using a common weather monitoring facility at the VLT site called Astronomical Site Monitor(ASM)~\cite{asm2000}. The measurements for seeing and PWV are done for zenith observations. So for the comparison, they are corrected for telescope pointing. 

\subsection{Science frames analysis}
The data reduction for all nights followed a common strategy, this include a chop subtraction: since chopping was performed between $\alpha$~Cen A and B, with either of them being on-axis in a single exposure. The chop subtraction was done in a way that the chopped frames showed $\alpha$~Cen A-B as coronagraphic PSF, and off-axis PSFs: $\alpha$~Cen A as negative and B as positive as shown in the Figure~\ref{fig:det}~(b). 

For frame selection, telemetry was calculated to remove the bad frames based on the criteria of AO correction, coronagraphic leakage, and background variance. An additional parameter, using the positions of the off-axis PSFs was used to co-align the frames.

The background variance was calculated, using a 10~pix square box at the four corners of the chopped frames, free from stellar residuals. The variance was measured for each box and then an average was calculated.

The coronagraphic residuals were measured using a 20~pix radius circular aperture from chopped frames. The position of the coronagraph was estimated using the position of the off-axis $\alpha$ Cen A/B.

Once good frames were identified, they were binned by averaging 500 frames. A rolling average approach was used to avoid adding of frames in case they were apart in the parallactic angle. The averaged frames showed a low spatial structure (residual sky-background, left-over after chopping), this was removed by applying a median filter of kernel size of 15. An example of such filtering is shown by Figure~\ref{fig:det}~(b).

\begin{figure}[H]
	\begin{center}
		\begin{tabular}{c} 
			\includegraphics[width=\textwidth]{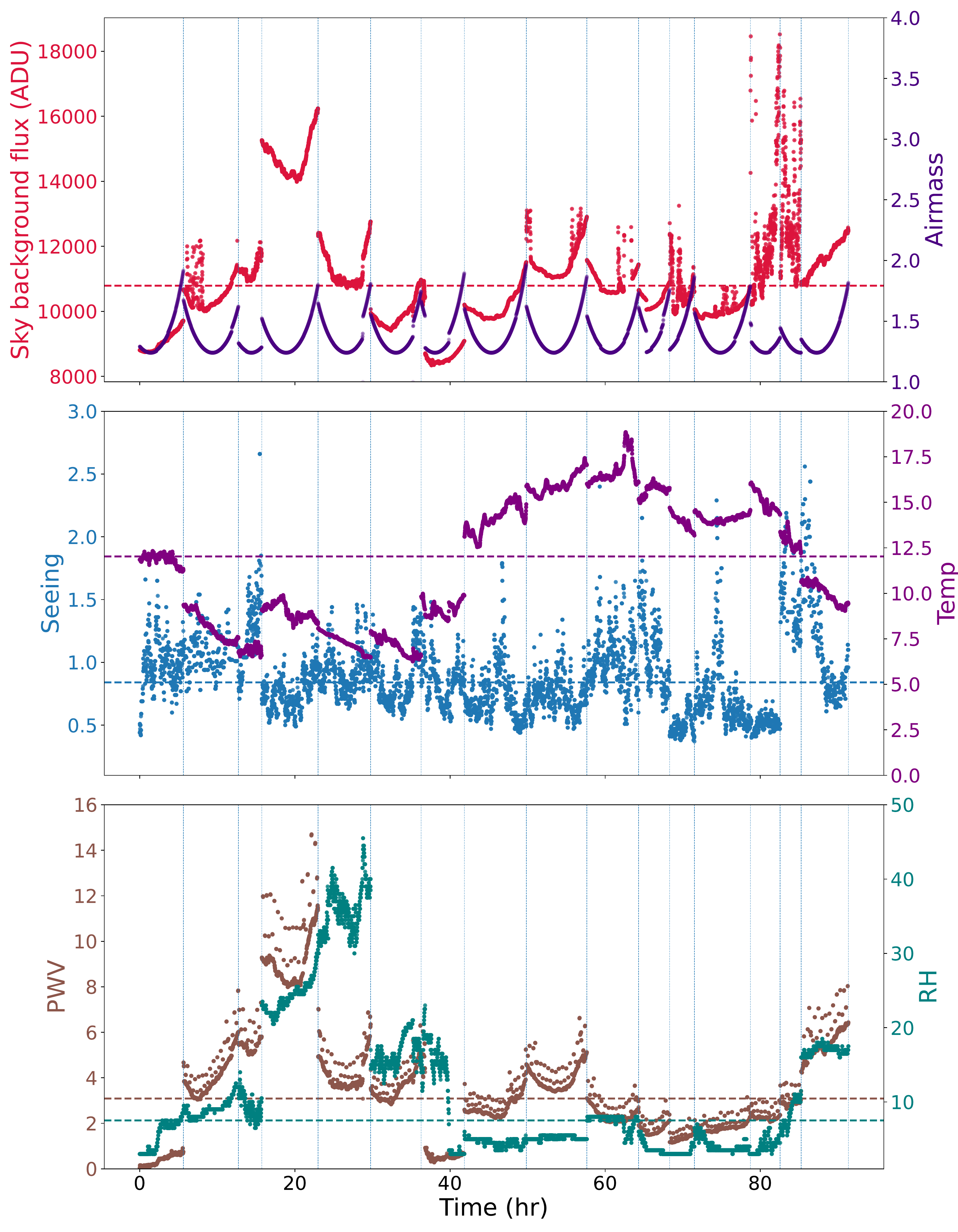}
		\end{tabular}
	\end{center}
	\caption[example] 
	{ \label{fig:sky_atm} 
		Thermal sky background compared with various atmospheric parameters such as temperature, PWV (mm), RM ($\%$), seeing ("), and effect of airmass. Vertical dashed lines separate different nights of observations and horizontal dashed lines represent median values.}
\end{figure} 
\begin{figure}[H]
	\begin{center}
		\begin{tabular}{c} 
			\includegraphics[width=\textwidth]{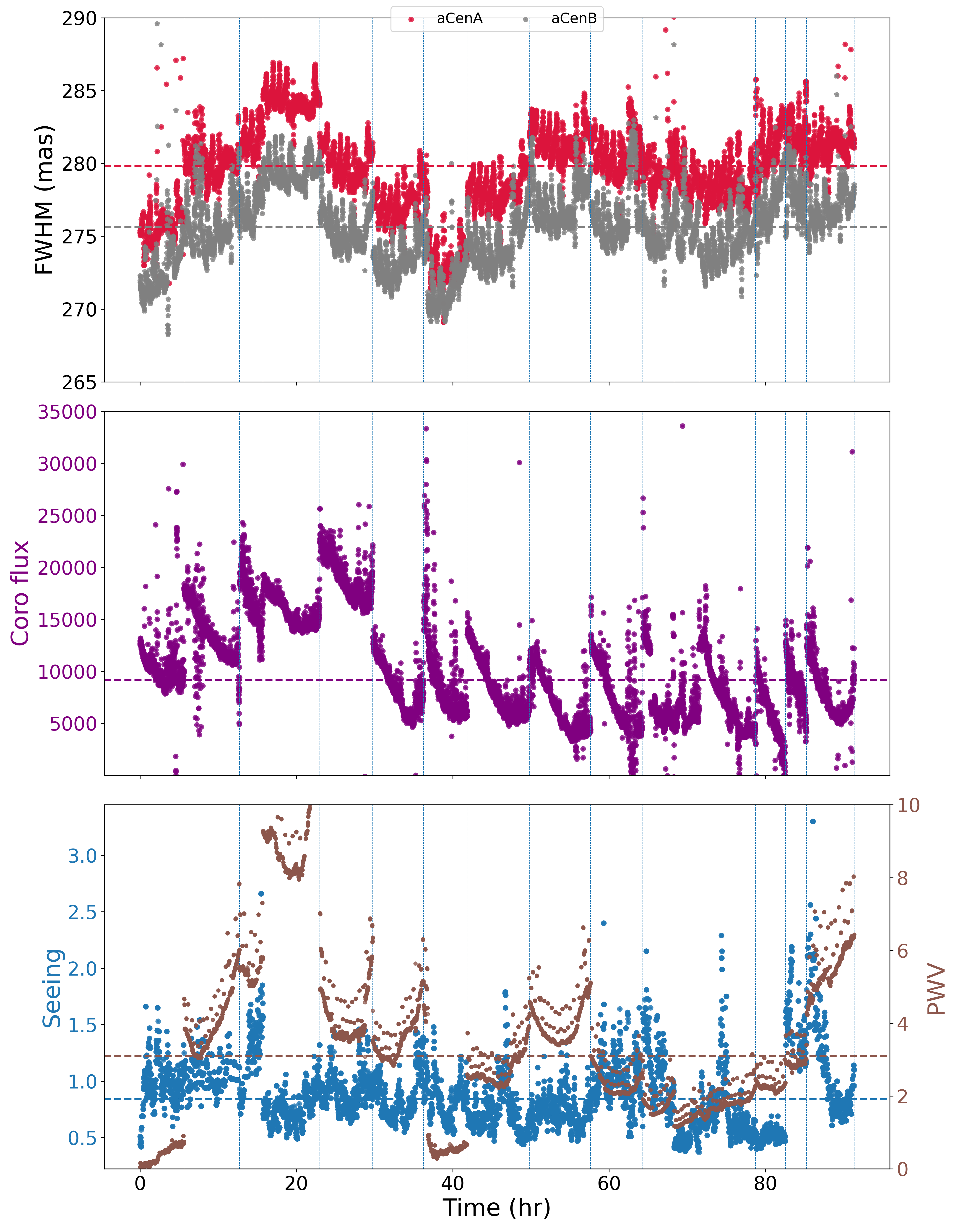}
		\end{tabular}
	\end{center}
	\caption[example] 
	{ \label{fig:AO} 
		Science PSF quality for $\alpha$ Cen A/B and coronagraphic performance compared with visible seeing and PWV.}
\end{figure} 
\begin{figure}[ht]
	\centering
	\begin{subfigure}[b]{0.49\textwidth}
		\centering
		\includegraphics[width=\textwidth]{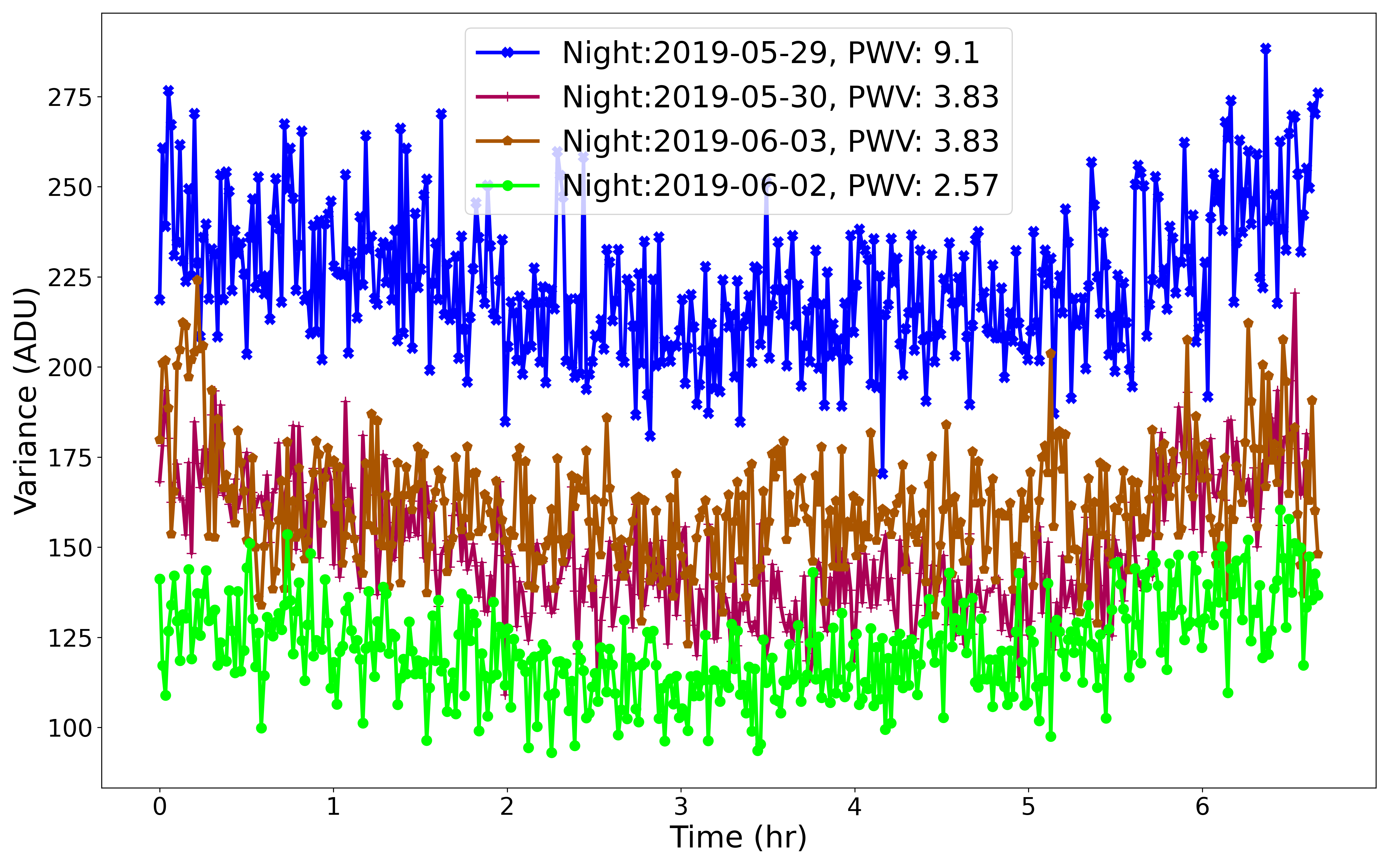}
		\caption{Sky background variance for different PWV values}
		\label{fig:variance}
	\end{subfigure}
	\hfill
	\begin{subfigure}[b]{0.49\textwidth}
		\centering
		\includegraphics[width=\textwidth]{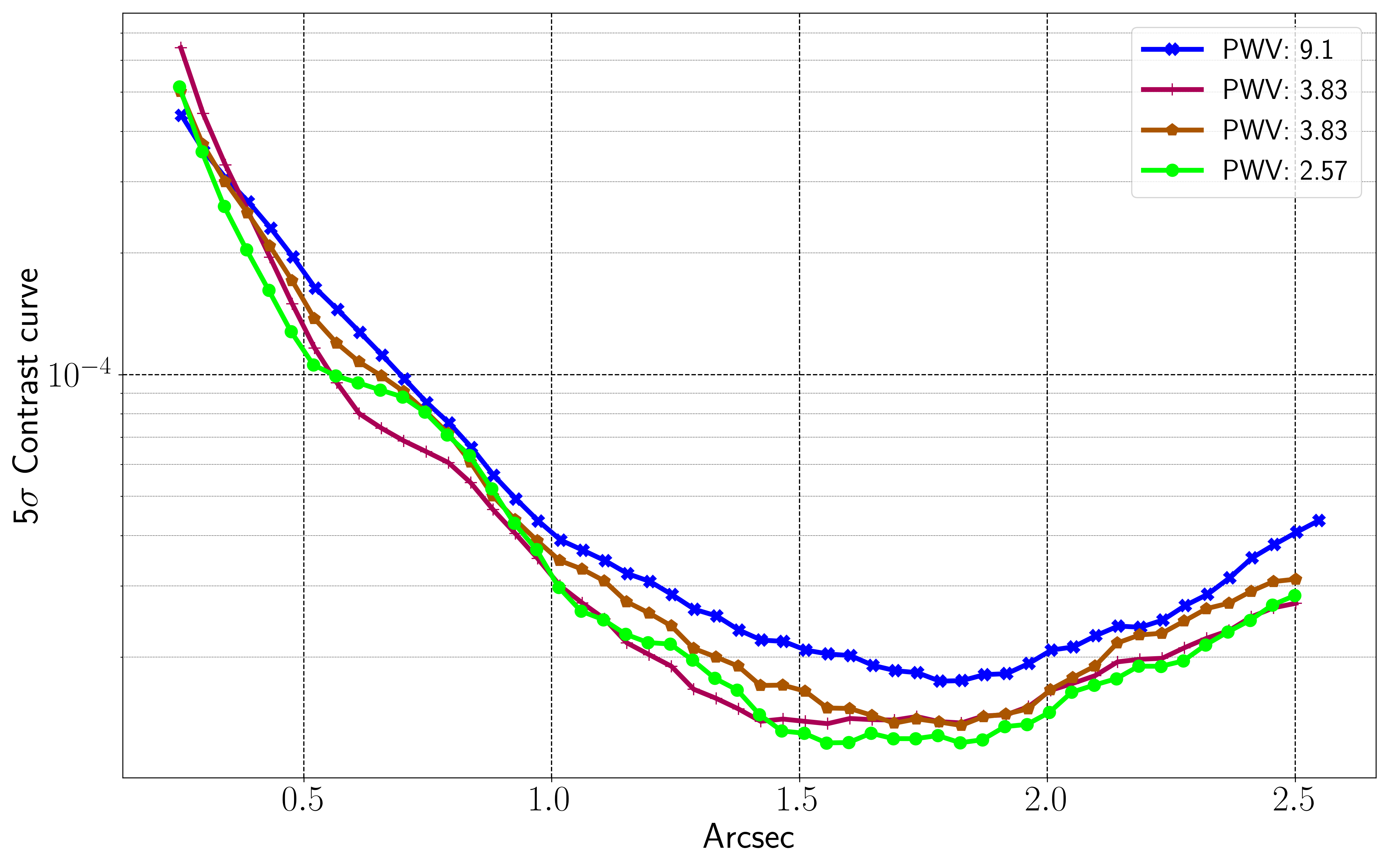}
		\caption{$5~\sigma$ contrast curves for different PWV values}
		\label{fig:cc}
	\end{subfigure}
	\caption{HCI performance under varying conditions of PWV. Four nights were selected with similar parallactic angles, airmass, and duration having different PWV values.}
	\label{fig:hci}
\end{figure}
\subsection{ADI analysis}
To study the impact of thermal sky background on the HCI performance, nights with similar parallactic angles and airmass were selected. This resulted in 9 nights out of 16, from these 9 nights, 4 nights were selected with low, high, and same PWV of 2.57, 9.1, and 3.83 respectively.

For ADI analysis, a full frame PCA routine based on Vortex Image Processing library~\cite{vip2017} with 10 principle components was used to process the data. 

\section{Instrumental limitations}
One of the limitations affecting the high-performance coronagraph at smaller separations was AGPM glow as shown in Figure~\ref{fig:det} (a). In future mid-IR instruments, this could be removed by incorporating a cold pupil stop in front of the AGPM mask. 

The other limitations come from the science detector. The detector has vertical read-out channels with different bias as shown in Figure~\ref{fig:det} (a). With chopping the effect of different channels is removed. But when a target passes through one channel to another, a charge leakage affects the photometry. Which renders the high precision photometric measurements. Persistence is another big limitation with such a detector as shown in Figure~\ref{fig:det} (c). The figure shows PCA processed image for one night of observation. The persistence stripe due to chopping between $\alpha$~Cen A/B is clearly visible. This limits the search area for the exoplanets. The faint arcs present in the image are part of off-axis PSFs.  

Development of mercury cadmium telluride (HgCdTe) based mid-IR detector arrays with lower-noise performance, shows a promising future for HCI instruments working in the mid-IR~\cite{geosnap}.

\section{Results}\label{sec:results} 
\subsection{Effect of atmospheric parameters}
A comparison between thermal sky background and various atmospheric parameters such as temperature, seeing, PWV and RH, as well as the effect of airmass is shown in Figure~\ref{fig:sky_atm}. Thermal sky background values presented in the figure are filtered for clouds, using a sigma clipping of 3.5 using astropy library. A strong correlation between sky background and PWV can be seen, and a weak correlation with RH exists. As expected sky background follows airmass. We find no correlation between visible seeing and temperature. In a previous work done by~\cite[Turchi et al]{turchi2020}, PWV is shown to have a direct impact on sky background IR emission in the $[10-12.5]~\mu m$ wavelength window. 

A comparison between science PSF quality and coronagraphic performance with visible seeing and PWV is shown in Figure~\ref{fig:AO}. The quality of AO correction in the science band is represented by FWHM of $\alpha$~Cen A/B. A strong correlation between science PSFs ($\alpha$~Cen A/B) quality and PWV is observed. We see visible seeing has no effect on the PSF quality, this shows that thermal background is the dominating factor in the N-band. 

For coronagraphic performance, no correlation with PWV, seeing, and PSF quality is seen. An incremental degrading coronagraphic performance can be seen for the first 5 nights of observation~\ref{fig:AO} (middle plot). After investigation, it was found that there was an ice formation on the AGPM coronagraphic mask. By incorporating small warm-up cycles the effect of ice formation was reduced. This is evident in the reduction of residuals for the rest of the campaign. 

The effect of PWV on sky background is shown by variance in Figure~\ref{fig:variance}. The background variance gets doubled by increasing PWV values from 2.57 to 9.1. The effect of increased sky background variance on high contrast performance is shown the Figure~\ref{fig:cc}. The figure shows $5~\sigma$ contrast curves for different values of PWV. The contrast curve gets degraded by $\approx50\%$ from PWV values of 2.57 to 9.1. In the presence of high PWV, the HCI performance is significantly degraded. 

\section{Summary and Conclusions}
In this work, we explore the effect of various atmospheric parameters and instrumental limitations on HCI performance in the N-band. We show that thermal sky background is one of the biggest limiting factors for the HCI observations in the mid-IR regime. The amount of thermal sky background is directly correlated with PWV. A high PWV can double background noise variance, which results in a degradation of contrast by $50\%$. 
%


\section{Acknowledgment}
The authors would like to thank the ESO and the Breakthrough Foundation and all the people involved for making the NEAR project possible. Part of this work has received funding from the European Research Council (ERC) under the European Union’s Horizon 2020 research and innovation programme (grant agreement No. 819155), and by the Wallonia-Brussels Federation (grant for Concerted Research Actions).

\bibliography{report} 
\bibliographystyle{spiebib} 

\end{document}